\documentclass[preprint,12pt,authoryear]{elsarticle}

\usepackage{amssymb}
\usepackage{amsmath}
\usepackage{lineno}

\usepackage{comment}
\usepackage{times}
\usepackage{graphicx}
\usepackage{booktabs}
\usepackage{natbib}
\usepackage{float}
\usepackage{perpage}
\usepackage[table]{xcolor}

\MakePerPage{footnote} 

\journal{Planetary and Space Science}

\begin{document}

\begin{frontmatter}

\title{Photothermal Spectroscopy for Planetary Sciences: Mid-IR Absorption Made Easy}

\author[label1,label2]{Christopher Cox} 
\author[label1]{Jakob Haynes} 
\author[label1,label2]{Christopher Duffey} 
\author[label2]{Christopher Bennett} 
\author[label1]{Julie Brisset} 

\address[label1]{Florida Space Institute, University of Central Florida, 12354 Research Pkwy, Orlando FL-32826, USA}
\address[label2]{Department of Physics, University of Central Florida, 4111 Libra Dr, Orlando FL-32816, USA}

\begin{abstract}

The understanding of the formation and evolution of the solar system still has many unanswered questions. Formation of solids in the solar system, mineral and organic mixing, and planetary body creation are all topics of interest to the community. Studying these phenomena is often performed through observations, remote sensing, and in-situ analysis, but there are limitations to the methods. Limitations such as IR diffraction limits, spatial resolution issues, and spectral resolution issues can prevent detection of organics, detection and identification of cellular structures, and the disentangling of granular mixtures. Optical-PhotoThermal InfraRed (O-PTIR) spectroscopy is a relatively new method of spectroscopy currently used in fields other than planetary sciences. O-PTIR is a non-destructive, highly repeatable, and fast form of measurement capable of reducing these limitations. Using a dual laser system with an IR source tuned to the mid-IR wavelength we performed laboratory O-PTIR measurements to compare O-PTIR data to existing IR absorption data and laboratory FTIR measurements for planetary materials. We do this for the purpose of introducing O-PTIR to the planetary science community. The technique featured here would serve to better measurements of planetary bodies during in-situ analysis. We find that, unlike other fields where O-PTIR produces almost one-to-one measurements with IR absorption measurements of the same material, granular materials relevant to planetary science do not. However, we do find that the materials compared were significantly close and O-PTIR was still capable of identifying materials relevant to planetary science.

\end{abstract}


\end{frontmatter}

\section{Introduction}
\label{s:intro}

Optical-PhotoThermal InfraRed (O-PTIR) spectroscopy is a relatively new method of spectroscopy \citep{spadea2021analysis}. It is used in other fields, such as plastics \citep[e.g., ][]{barrett2020microplastic} and life sciences \citep[e.g., ][]{klementieva2020super}, but here we present this technique in the context planetary science.

Our specific interest lies in understanding the early formation of solids in the solar system, how minerals and organics came to be and were mixed, creating the planetary bodies we see today. While remote sensing allows us to determine the general composition of objects from afar \citep{chauhan2015hyperspectral}, we know from the study of meteorites, that minerals are often mixed at the sub-micron level \citep{brownlee2006comet}. This is true in particular in small undifferentiated bodies, such as asteroids and comets \citep{hanner2004composition,lasue2006porous,thebault2019there}, but also in the regolith covering larger bodies, such as the Moon \citep{thompson2010smallest,zbik2020possible} and Mars \citep{zeng2021revealing}. Detecting the compositional details at such high spatial resolutions allows us not only to learn the exact composition of primitive materials in the solar system but also to understand the conditions and processes of their formation. This is something only a close-up analysis can provide, be it on meteorites or returned samples, or in situ on planetary surfaces. Furthermore, it can only be achieved via non-destructive analysis methods, in order to preserve the compositional information as much as possible.

In addition to mineralogical studies, missions to distant bodies also have biological components (e.g. Viking Lander, Europa Clipper, etc.). Particularly, there is interest in studying prebiotic molecules and organics in the context of their environment and formation\footnote{Origins, Worlds, and Life: A Decadal Strategy for Planetary Science and Astrobiology 2023-2032}. The search for these organics is ongoing on meteorites \citep[e.g.][]{shimoyama1997complex,chan2022water}, as well as in-situ, on Mars, for example \citep{ansari2023detection}. Here too, high-spatial resolution, non-destructive analysis techniques are best suited for obtaining as much information as possible from these extraterrestrial samples. Since we expect organics to be mixed within primitive materials, micron to sub-micron resolutions are optimal for their detection and mapping. In addition, the sizes of living cells on Earth can range from the sub-micron level to multiple microns. Therefore, potentially detecting such a cell-like structure in extraterrestrial materials would also require high-spatial resolution analysis tools. 

Current non-destructive analysis techniques suffer from limitations on spatial resolution. In particular, organics and many minerals are best studied in the IR 
Since the diffraction limit of optical systems (e.g. FTIR) is on the order of the wavelength of the light it utilizes, it inhibits micron to sub-micron resolutions in those systems. In addition, many organic molecules are best detected in absorption. For this reason, sample vaporization followed by mass spectrometry is sometimes used for their detection \citep{kandiah2013advances}. While able to detect prebiotic building blocks \citep{sokol2011miniature}, such destructive techniques are sure to break apart any potential complex organic molecules, such as proteins, or cell-like structures.

In this context, we have looked into the possibility of leveraging the O-PTIR technique for studying the composition of extraterrestrial samples. This method detects absorption in the mid-IR and is capable of performing non-contact chemical analysis at a sub-micron level and in a non-destructive manner. In addition to individual measurement points, O-PTIR has the ability to scan areas including large gaps, such as those found in granular samples. This has been highlighted as an advantage by the lessons learned from the PIXL instrument on NASA's Mars rover, Perseverance \citep{tice2022alteration}. The fast measuring speed of O-PTIR (demonstrated and discussed later in this work, Section \ref{s:fast-meas}) allows for the possibility to scan such large sample areas, mapping their IR absorption spectra. 

\section{Methodology}
\label{s:methods}

\subsection{O-PTIR}
\label{s:optir-intro}

The O-PTIR technique is based on the principles of thermal lens spectroscopy which is well described by \cite{snook1995thermal}. O-PTIR is a pump-probe technique that utilizes the photothermal effect using two lasers: one in the IR (pumping laser) at the chosen frequency, and one in the visible (probing laser) to probe the thermal expansion of the sample \citep{bazin2022using,paulus2022correlative}. O-PTIR spectroscopy pushes past the diffraction limit of traditional IR techniques and provides a significant increase in spatial resolution \citep{bazin2022using,paulus2022correlative}. The IR laser is, typically, a Quantum Cascade Laser (QCL). A QCL is a semiconductor laser that emits in the infrared region of the EM spectrum. This allows for a range of IR frequencies that the QCL can access by varying energies in the laser. The IR absorption of the sample causes a photothermal effect on the target. This consists of thermal expansion of the surface of the sample and a change in the index of refraction of the sample \citep{olson2020simultaneous}. As the QCL scans across its frequency range, the photothermal effect changes on the surface allowing for a measurement at each frequency.

\begin{figure}
    \centering
    \includegraphics[width=\textwidth]{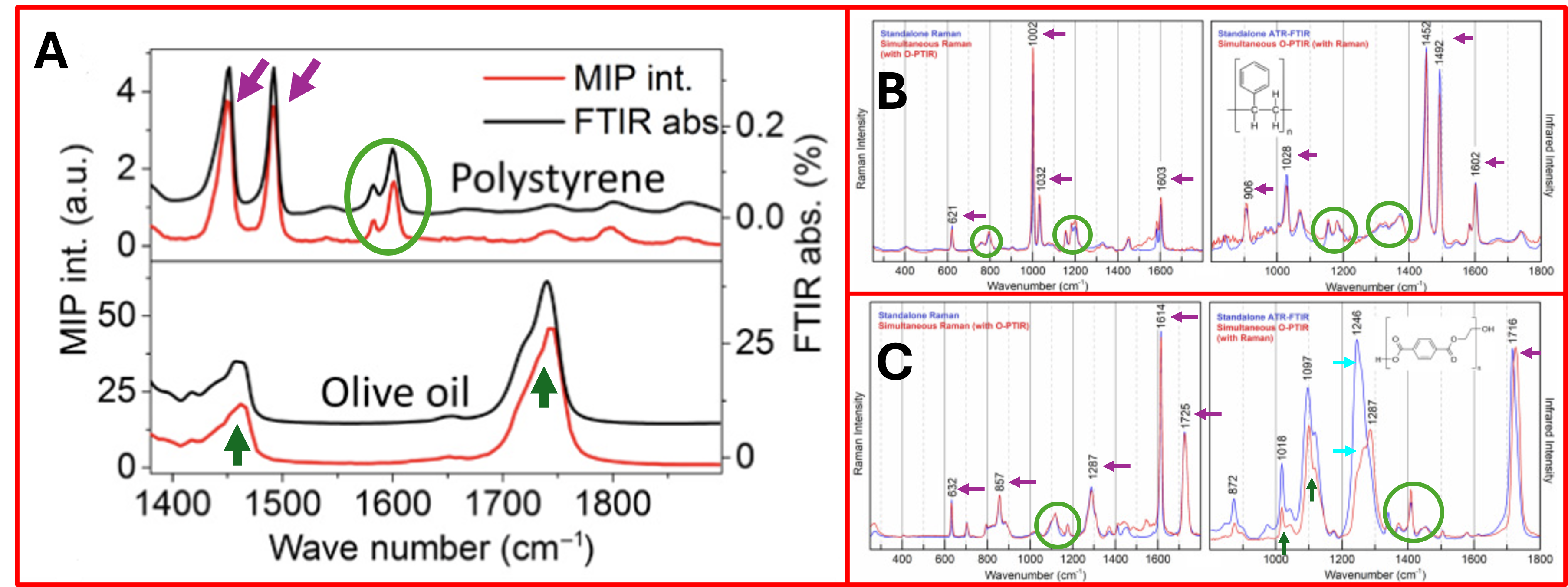}
    \caption{Demonstrations of the ability of O-PTIR to produce spectra similar to existing and popular methods. A.) MIP (red line) is a Mid-Infrared Photothermal measurement. The black line is a traditional FTIR-A measurement. Both the top and bottom are offset for clarity. The purple arrows superimposed on the polystyrene (top) plot point out the one-to-one capability of the O-PTIR method. The light green circle superimposed on the polystyrene plot shows matching features. The dark green arrows on the olive oil (bottom) plot show matching features with the peaks slightly offset. Original figure courtesy of \cite{zhang2016depth}. B.) O-PTIR with simultaneous Raman compared to standalone Raman (left) and standalone ATR (right) for polystyrene. The purple arrows show peaks matching one-to-one and the green circle shows matching features. Original figure courtesy of \cite{krafft2022optical}. C.) O-PTIR with simultaneous Raman compared to standalone Raman (left) and standalone ATR (right) for PET. The purple arrows show peaks matching one-to-one and the green circle shows matching features. The dark green arrows indicate corresponding peaks that are slightly shifted. The cyan arrows indicate a pair of peaks that are slightly offset and have differing spectral shapes. Original figure courtesy of \cite{krafft2022optical}.}
    \label{f:optir-demo}
\end{figure}

O-PTIR spectroscopy is capable of producing FTIR absorption-like spectra at a sub-micron spatial resolution \citep{bazin2022using}. That is to say, the features and peaks of O-PTIR spectroscopy are commonly similar to FTIR absorption spectra. This is demonstrated in Figure \ref{f:optir-demo}. O-PTIR is a non-destructive method already in use by the life science community \citep[][among others]{spadea2021analysis} and the microplastics community \citep[][among others]{chen2022portable}. These works provide many examples comparing this technique to FTIR absorbance and transmittance techniques to demonstrate its validity \citep[e.g.][]{bazin2022using,zhang2016depth,krafft2022optical}. Past works have found strong correlations between the methods. Previous works focus on solids and prepared sample slides. Our use of O-PTIR will focus on the study of granular materials and materials relevant to planetary science. 

The spatial resolution and capability of this technique allow for the compositional study of materials at the sub-micron level. This allows for the specific study of individual grains within a mixture. This is important due to the ability to further study individual components of a larger mixture such as what would be measured by an orbiting remote sensing spacecraft. Additionally, the sub-micron resolution has advantages in astrobiology as it would allow for analysis into possible cell-like structures on foreign bodies. O-PTIR can be coupled with simultaneous Raman measurements. This technique is demonstrated in some of the data presented in Figure \ref{f:optir-demo}. The addition of Raman to the O-PTIR technique allows for better and unambiguous identification of minerals and amino acids.

\subsection{Mid-IR Wavelength Range}

The wavenumber range that we are studying in the present work is 1800 - 980 cm$^{-1}$ (5.6 - 10.2~$\mu$m). This is due to the areas of interest within that range. The "fingerprint region,"
a region that contains many unique complex absorption features, lies in the 1500 - 500 cm$^{-1}$ (6.67 - 20.0~$\mu$m) range \citep{hu2016far}. The fingerprint region is key to unambiguously identifying minerals and varying compounds. Several key mineralogical features lie specifically in the 1300 - 805 cm$^{-1}$ (7.7 - 12.4~$\mu$m) range. Additionally, many functional groups (e.g., O-H stretches, C-H stretches, C=O stretches) lie in the region of 3600 - 1500 cm$^{-1}$ (2.78 - 6.67~$\mu$m), which is excellent for the detection of organic species \citep{schmitt1998ftir}. O-PTIR spectra are comparable to absorption spectra \citep{bazin2022using}. This means this technique is capable of identification and qualification of components. This is an improvement of the spectra typically measured on planetary surfaces. In this paper, we will describe the sample preparation, how measurements were taken, and analysis of the data produced in Section \ref{s:methods}. In Section \ref{s:characterization} we will describe the instrument used in this measurement campaign as well as offer characterizations for the instrument. In Section \ref{s:discuss} we will discuss the benefits of the instrument and technique for planetary science. Finally, in Section \ref{s:conclusion}, we will conclude the paper and summarize our work.

\subsection{The mIRage\textsuperscript{\textregistered} Instrument}
\label{s:mirage-intro} 

The mIRage\textsuperscript{\textregistered} instrument is an O-PTIR instrument that utilizes a probing visible laser (0.532 $\mu$m) and a tunable pulsing (pumping) IR laser (5.494 - 10.416 $\mu$m; 1820 – 960 cm$^{-1}$). The visible laser is a 200 mW 
laser. 
The IR laser is a 300 mW laser. 
mIRage\textsuperscript{\textregistered} produces a spectral measurement at each wavenumber the IR laser is tunable to, so it has the capability to produce a spectral resolution of 1 cm$^{-1}$ and almost 900 spectral channels. The number of spectral channels is dependent on the IR source. The spectral range is set by what the IR source is tunable to.

The spectral resolution is also controlled by the IR source and the settings on the IR source. Our current understanding of the instrument and technique also allows for varying spectral resolutions. The mIRage instrument allows for a sacrifice of resolution to increase the speed of a measurement and vice-versa. 

\begin{figure}
    \centering
    \includegraphics[width = \textwidth]{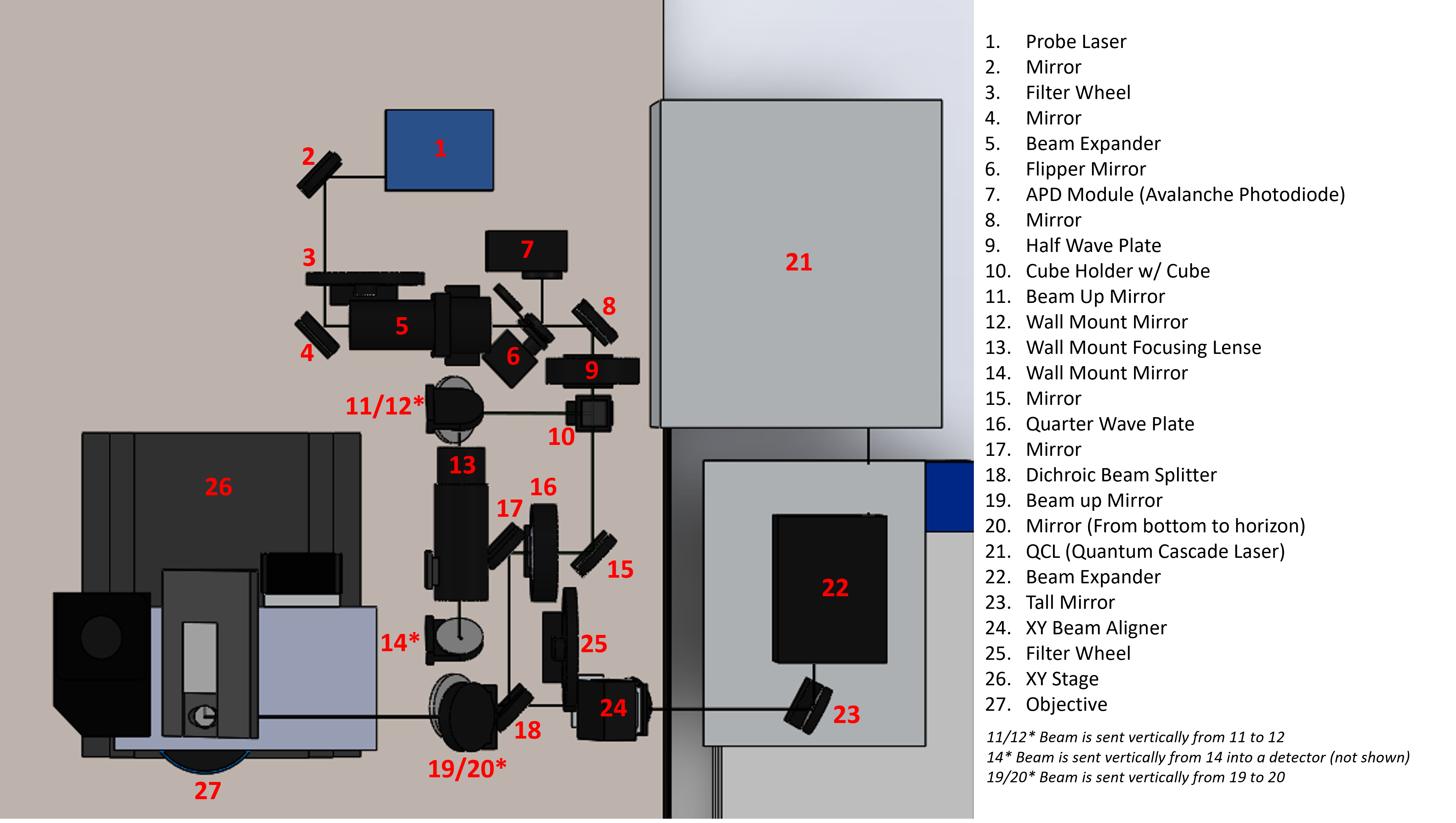}
    \caption{Image of the optical path of the mIRage\textsuperscript{\textregistered} instrument. Infrared and visible lasers are directed to the same area of a surface for measurement using the O-PTIR technique. The optical laser is directed back from the sample to the detector and a measurement can be taken. The individual parts are labeled in the figure.}
    \label{f:method-concept}
\end{figure}

The optical path is depicted in Figure \ref{f:method-concept}. Both lasers are variable power lasers. The instrument is designed and produced by Photothermal Spectroscopy Corp\footnote{https://www.photothermal.com/}.

The sample is viewed using the instrument's optical microscope and objective lenses. There are two objectives with varied magnification. The low objective has less magnification than the high objective and is used to find points of interest in a sample. When a point of interest is found, the microscope is switched to the high objective. The high objective has more magnification than the low objective and is used to focus on the point of interest. The dual laser system passes through the high objective. The instrument produces measurements by emitting a visible laser into the distortion (photothermal effect) created by the IR laser. This photothermal effect is better described in Section \ref{s:optir-intro}. The visible laser reflects back into the detector and the difference of intensities compared to incoming light is what produces the O-PTIR spectra that will be featured in Section \ref{s:characterization}. In the case of automated multiple measurements (i.e. a hyperspectral map), the instrument performs a quick and automated auto-focus prior to each individual measurement.

\subsubsection{Instrument Calibration}
\label{s:calibration}

Every time the instrument is turned on, the instrument is initialized and calibrated. The calibration process uses a known and well-accepted standard (MirrIR low-e microscope slide\footnote{http://kevley.com/MirrIR.aspx}, Kevley Technologies\footnote{http://kevley.com/About.aspx}). A spectrum and two single-frequency heat maps are taken of the standard sample. The two heat maps are used to calibrate both QCL chips that compose the IR laser, so each heat map uses one frequency from each QCL chip. The heat map is a measure of intensity at a specific location of the sample at a specific wavenumber. This baseline spectrum is used to show the intensity of features of the noise in the system. The instrument software subtracts this background from each measurement. Humidity of the sample area, optimization of the alignment, and auto-focusing are also measured and calibrated as part of the calibration process. Background and calibration results are typically consistent but can change with varying humidity. Humidity was constantly monitored (whether spectra were being collected or not) and if the humidity changed by more than 5\% from the previous calibration, the instrument was re-calibrated by a user.

Though the O-PTIR method of measuring is non-destructive, both lasers in the dual laser system are capable of damaging a sample. The IR laser is capable of bleaching the surface of a sample. The probe (visible) laser has enough power to blow a hole in the sample which can cause a sort of secondary cratering from that destruction. The calibration process and part of the measuring process (described in Section \ref{s:measurements}) are utilized the avoid this damage to the sample.

\subsection{Sample Preparation}
\label{s:sample-prep}

\begin{figure}[H]
    \centering
    \includegraphics[width=\textwidth]{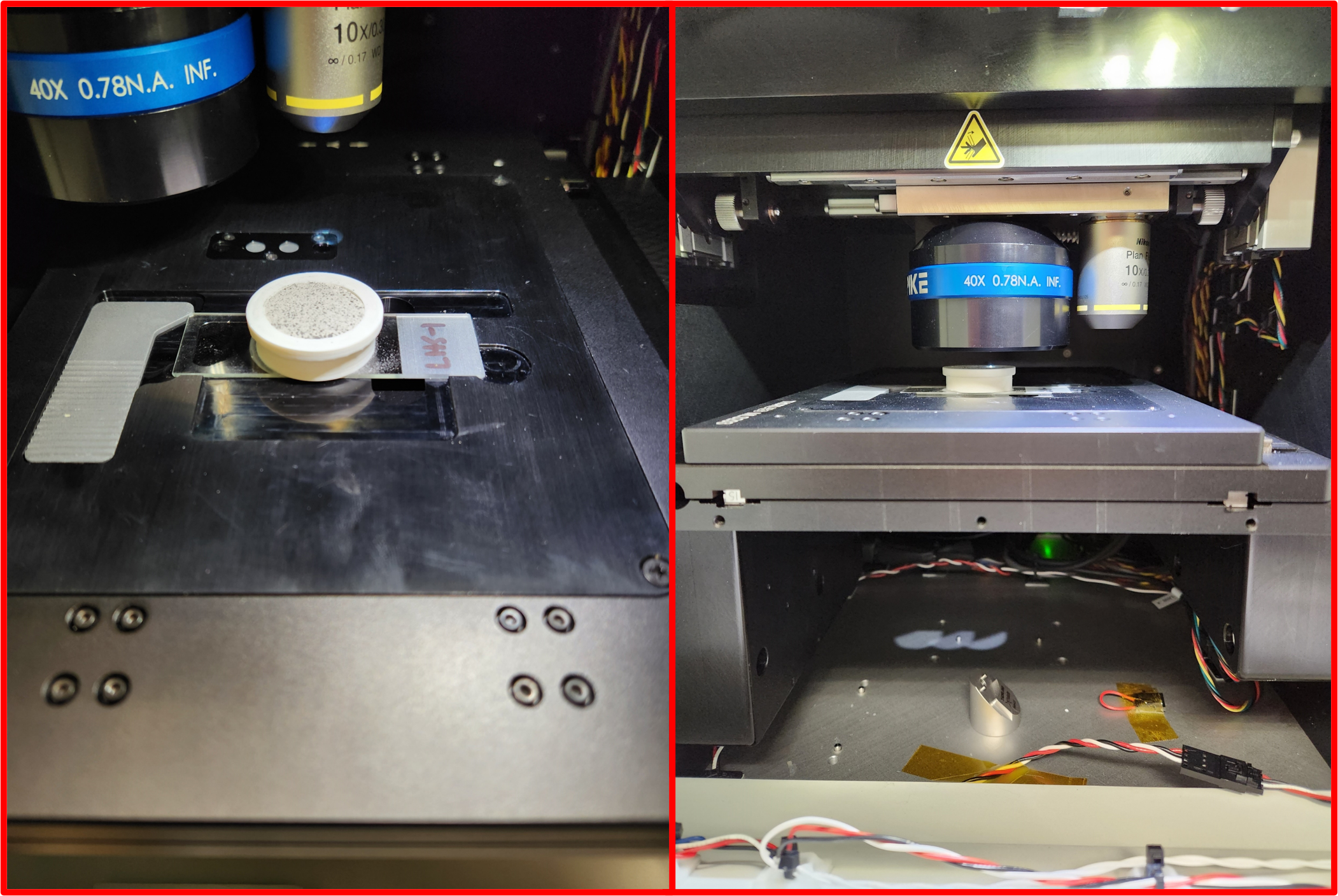}
    \caption{Image of the sample holder (left) and sample area of the mIRage\textsuperscript{\textregistered} instrument (right). The sample holder is a 3D-printed cylinder adhered to a glass slide that fits into the sample area. The sample area is an optical stage designed to move (in three dimensions) the sample under the instrument's microscope.}
    \label{f:sample}
\end{figure}

Though specific sample preparation is not required by the technique, we describe here how we prepared the samples for consistency in measurements and cleanliness of the instrument. Cylindrical sample holders were 3D printed and adhered to glass slides that fit into the mIRage\textsuperscript{\textregistered} instrument's sample holder. This sample holder can be seen in Figure \ref{f:sample}. The granular material was poured into a sample holder and flattened to reduce the height variation at the surface. The flattening process involved a light packing into the sample holder and then a straight edge was scraped over the surface of the sample holder. The packing was to prevent granules from shifting during transport and the straight edge was used to ensure a flat surface as well as allowing the container to be sealed. The sealing of the sample container is solely for the purposes of protection, storage, and transportation.

\subsection{Measurements}
\label{s:measurements}

A measurement begins with the loading of a sample into the sample area (see Figure \ref{f:sample}) and after calibration is completed. The objectives are moved to be over the sample. To avoid damage and destruction of the sample, the power of the IR laser and visible laser are both adjusted per material. This information is recorded per spectrum. The power levels are adjusted until there is sufficient signal being returned to the sensors for both lasers. The higher the power in both lasers, the better the signal-to-noise ratio. Gain can also be used to increase the signal without increasing power and damaging the sample. Gain was changed and recorded per material.

A single spectral measurement is taken of a single grain of a sample. A single grain is focused on using the optical cameras and the high objective of the instrument. For the purposes of this work, a single spectral measurement will be a set of three spectra taken at a single point on the sample that are averaged together. The software for the instrument averages this set of spectra in real time into one spectrum. Multiple single spectral measurements are taken across an area of a sample to make sure no damage is being done to any part of a sample as this risk changes for different materials (for mixtures) within a sample. Once the safety of the sample is ensured, a hyperspectral map is made of the sample. A hyperspectral map is, in this case, over 1000 individual hyperspectral measurements across the area of a sample. This is done to scan multiple grains to average out variations due to grain orientation, an issue demonstrated by \cite{jaret2018microspectroscopic,martin2018investigating,pernet2017assessing}, and to include multiple materials for mixtures. The instrument reports the spectral resolution of these maps as being 4.

\subsubsection{Hyperspectral and Heat Maps}
\label{s:hs-maps}

\begin{figure}[H]
    \centering
    \includegraphics[width = \textwidth]{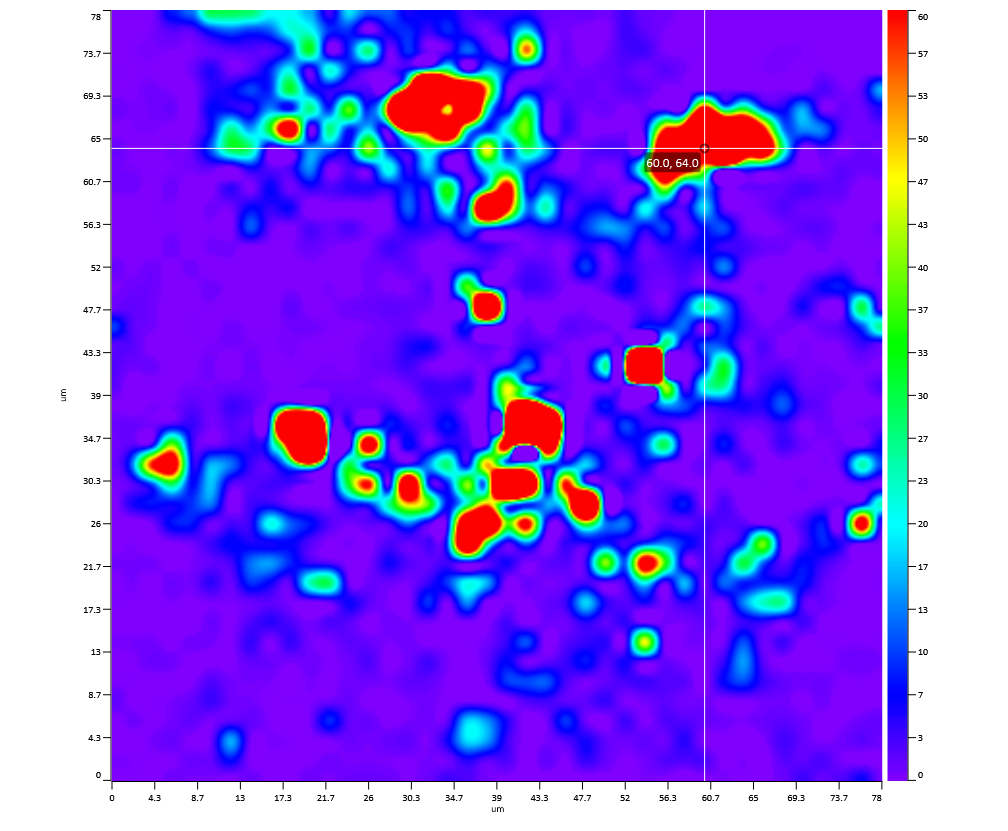}
    \caption{Example image of a single heat map in the stack that makes up the hyperspectral map used in this paper. This map (78 x 78 $\mu m$, 2 $\mu m$ spacing) is of bronzite and has a total of 1600 total hyperspectral points. The map took approximately 7 hours and 21 minutes to complete. The intensity of the signal is set to reflect the intensity of the various points in the map at 1017 cm$^{-1}$. The red areas are areas of most intense signal ranging down to purple being the least intense or no signal at the given wavenumber.}
    \label{f:hs-map-example}
\end{figure}

Figure \ref{f:hs-map-example} shows an example of a slice of a hyperspectral map, represented as a contour plot, generated by the O-PTIR technique. A hyperspectral map is a grid (in this case, evenly spaced) of hyperspectral measurements. For each point, a complete spectrum is produced for the wavelengths we are considering before the instrument moves the optical lens to the next location. The map that is produced from the grid of measurements is a sort of "heat map" where the map reflects intensity at a given signal. This is to say that a heat map is produced for each frequency measured. The stack of heat maps reflects the intensity of the sample at multiple hundreds of wavenumbers.

\begin{figure}[H]
    \centering
    \includegraphics[width=\textwidth]{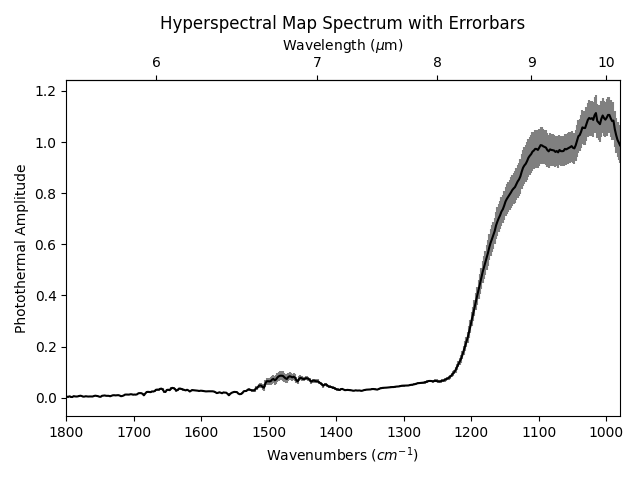}
    \caption{Plot of the hyperspectral map spectrum of anorthosite plotted with the standard error of the mean error bars. For the O-PTIR measurement, instrument settings were: 46\% IR Power, 3.5\% Probe Power, and 10x Detector Gain. The O-PTIR measurement is an average of a hyperspectral map (612 x 468 $\mu m$, 12 $\mu m$ spacing) which contained 2080 individual spectra.}
    \label{f:hs-error}
\end{figure}

Thresholds were set to distinguish peaks from noise in the measurement. Figure \ref{f:hs-error} shows a constituent material (anorthosite) plotted with the measurement and error bars. The error bars are the standard error of the mean over 2080 individual points. In this context, each point has almost 900 wavenumbers. With the error bars, the peaks are still visible and stand out against the noise. 

\subsection{Comparable Forms of Spectroscopy}
\label{s:spectroscopy-intro}

Currently available IR spectroscopy techniques for planetary sciences include Fourier Transform InfraRed (FTIR) and Attenuated Total Reflectance (ATR, also known as ATR-FTIR). FTIR can be best compared to O-PTIR. FTIR instruments are typically divided into two categories, absorbance (the more comparable of the two) and reflectance. Absorbance is the measure of a material's ability to absorb light and reflectance is the measure of the proportion of light that is reflected off the surface of a material. These measurements and their relationships will be discussed further below.

FTIR is a non-destructive form of chemical characterization of geological samples \citep[e.g.][]{griffiths1983fourier,matteson1993quantitative,chen2015applications}. Sometimes the sample preparation can be destructive due to the involvement of crushing and sieving materials in the sample preparation process, but the method itself is non-destructive \citep{farmer1974ism}. In vibrational spectroscopy, FTIR works by measuring transitions of quantized vibrational energy states. Absorption of IR radiation when a molecule is excited to a higher energy state causes additional vibrations of molecular bonds which happens at varying wavelengths in the measuring region (mid-IR in the case of this work) \citep{griffiths1983fourier}. This method (FTIR Absorption) requires light to pass through a sample. The optical properties of certain materials require the sample to be extremely thin ($< 50 \mu m$). The thin material is typically accomplished using a pellet method which produces a "pellet" (also known as a "disc" or a "wafer") of material thin enough for some amount of light to pass through on the way to a detector \citep{farmer1974ism}.

FTIR Reflectance (FTIR-R) techniques work similarly to FTIR Absorbance (FTIR-A) techniques \citep{chen2015applications}. Unlike absorbance techniques, reflectance techniques do not require light to pass through the entirety of the sample \citep{soares2014introduction}. Exceptions exist, but in these cases, FTIR-R does not provide chemical information about an entire sample \citep{mustard2019theory}. However, this method (FTIR Reflectance) does still provide chemical information on functional groups distributed near the surface \citep{chen2015applications}. 

ATR is a specific and common form of FTIR. Though this is considered a reflectance technique, ATR produces spectra closer to that of FTIR-A techniques due to its interaction with the sample. There are several benefits of this technique including reduced sample preparation and the ability to increase signal-to-noise ratios (SNR) \citep{ramer2006attenuated}. Particularly, SNR can be increased by increasing the signal's interaction with the sample (i.e. the number of bounces/reflections off the sample) \citep{ramer2006attenuated}.

Minerals are often well characterized in IR wavelengths and specifically in the wavelengths of interest for this work. Examples of this include multiple openly available databases. Our work utilizes data from Infrared and Raman Users Group (IRUG)\footnote{http://www.irug.org/search-spectral-database} and SpectraBase\footnote{https://spectrabase.com/}. These databases were selected for their open access and the availability of data for materials relevant to planetary science. Data from IRUG and SpectraBase are absorbance spectra. Additionally, several minerals are discussed and described in \textit{The Infrared Spectra of Minerals} \citep{farmer1974ism}.

\begin{figure}[H]
    \centering
    \includegraphics[width=\textwidth]{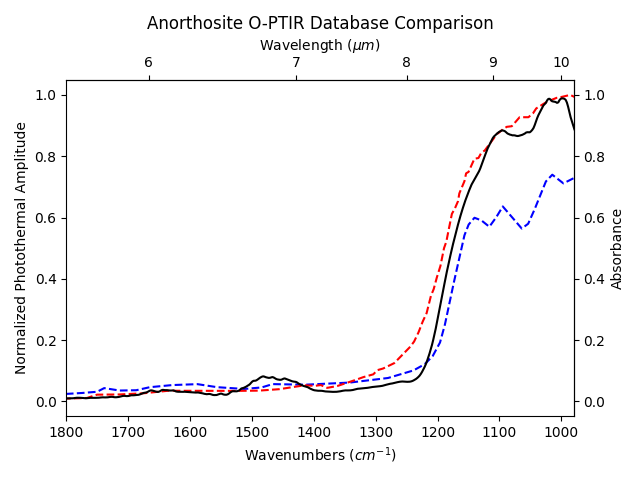}
    \caption{Mid-IR measurements of anorthosite and anorthite. The solid black line is the measurement produced using the O-PTIR technique. Absorbance data for comparison obtained from IRUG (blue, dashed) (Entry: IMP00197) and SpectraBase (red, dashed) (Spectrum ID: HGmXKwVLmf). In this case, the IRUG and SpectraBase data are anorthite (a plagioclase mineral). For the O-PTIR measurement, instrument settings were: 46\% IR Power, 3.5\% Probe Power, and 10x Detector Gain (see \ref{s:mirage-intro} for details). The O-PTIR was normalized by a factor of 0.92 (i.e. every point in the spectrum was multiplied by 0.92) to be more visibly comparable to absorbance database entries. The O-PTIR measurement is an average of a hyperspectral map (612 x 468 $\mu m$, 12 $\mu m$ spacing) which contained 2080 individual spectra.}
    \label{f:database-comparison}
\end{figure}

Figure \ref{f:database-comparison} shows our own O-PTIR measurement compared to FTIR database data (IRUG and Spectrabase). The similar spectral shapes and matching peak locations show consistency between the O-PTIR measurement and the FTIR-A data. The consistency between measurements demonstrates O-PTIR's ability to compare to existing data and other spectroscopy methods (FTIR-A). Anorthite was selected as the comparison mineral because we did not have database absorbance data of of anorthosite to compare to. However, anorthite is a plagioclase mineral and anorthosite is a rock composed of other plagioclase minerals. According to the source for the anorthosite sample, our anorthosite sample is \~90\% anorthite. As it has been noted above, O-PTIR is comparable to absorption spectra. An exact match was not expected given the anorthosite sample containing up to 10\% minerals other than anorthite. It can be seen that the O-PTIR data shares spectral features and spectral shapes with the IRUG absorbance spectrum. Additionally, there are similar and matching peaks with both the IRUG and SpectraBase spectra presented here. The two database entries could be inconsistent for reasons including grain orientation and grain size distributions. That information was unavailable in both database entries. Further, \cite{farmer1974ism} lists peaks associated with anorthite in the IR. In the examined wavenumber range, anorthite peaks at 1020, 1085, and 1160 cm$^{-1}$. The peaks at 1020 and 1085 cm$^{-1}$ correspond closely with two of the measured O-PTIR peaks.

\subsubsection{FTIR Comparisons}
\label{s:ftir-methods}

In addition to the O-PTIR measurements, which we are presenting here, we performed a few FTIR measurements to serve as a comparison. This comparison is intended to show the similarities and differences in measurements (O-PTIR vs. FTIR) as well as display the advantages of O-PTIR over FTIR. We present those in more detail in Section~\ref{s:ftir-comp}. Here, we describe our methodology for these FTIR measurements. 

Each of our FTIR measurements utilized a pure KBr pellet as a background and an anorthosite sample pellet as comparable measurements. The anorthosite pellet was a 200.9 mg pellet consisting of 198.4 mg KBr and 2.5 mg anorthosite. This yielded a 1.24\% anorthosite pellet. The background and the sample were scanned in the same way. Each pellet was measured using an iS50R FTIR with the OMNIC software used to collect the spectra. Each measurement was taken with a gain of 2.0 and an aperture of 87. Each pellet was measured with multiple scan settings (1, 3, 5, 8, 16, 32, 64, 128) at multiple resolutions (2, 4, 8, 16). Each measurement setting was performed 3 times.

\section{O-PTIR Characterization for Planetary Science Applications}
\label{s:characterization}

\subsection{Repeatability}
\label{s:repeatability}

\begin{figure}[H]
    \centering
    \includegraphics[width = \textwidth]{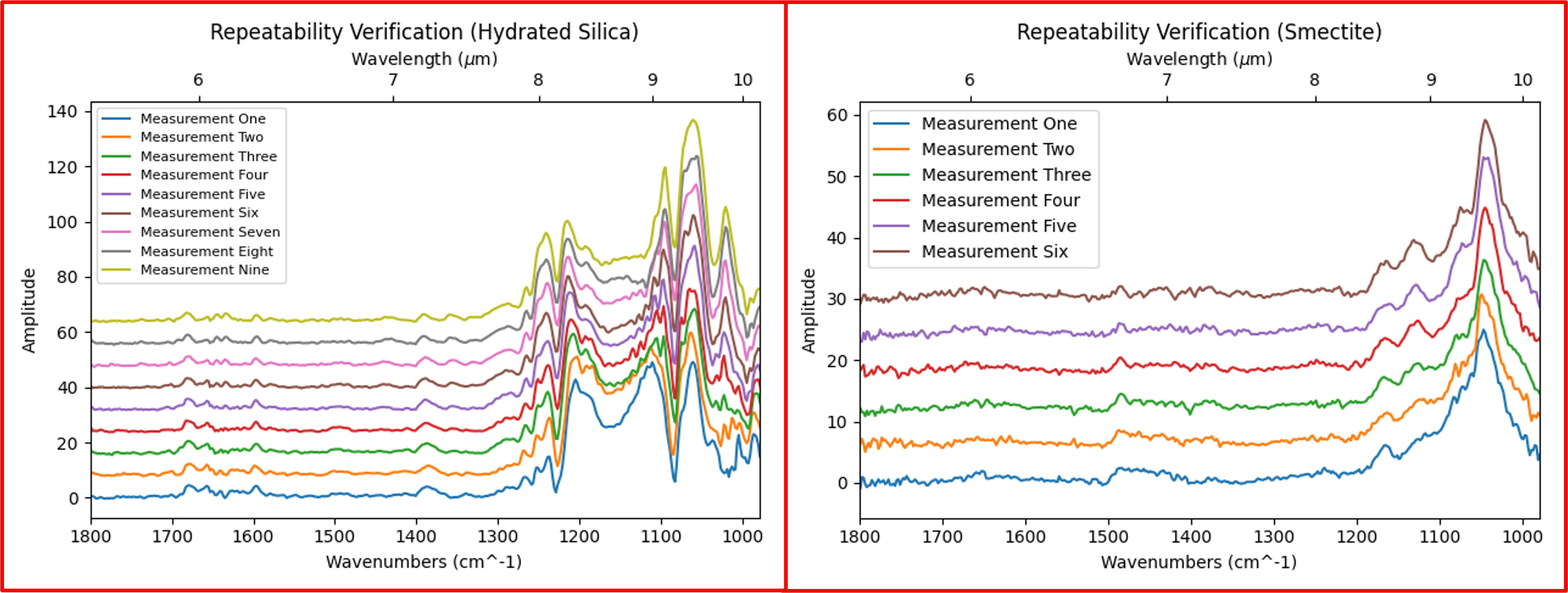}
    \caption{Mid-IR O-PTIR measurements of hydrated silica (left) and smectite (right). The measurements presented here are individual spectral measurements. For the hydrated silica, instrument settings were: 46\% IR Power, 3.5\% Probe Power, and 5x Detector Gain. There were nine measurements of hydrated silica. For the smectite, instrument settings were: 21\% IR Power, 2\% Probe Power, and 10x Detector Gain. There were six measurements of smectite. The hydrated silica measurements are offset by 8 units and the smectite measurements are offset by 6 units for ease of readability.}
    \label{f:repeatability}
\end{figure}

Figure \ref{f:repeatability} shows measurements of hydrated silica (left) and smectite (right). Each set of measurements used the same power settings and location. These measurements were used to demonstrate the repeatability of the instrument. For each respective set of measurements, the spectra peak at the same locations, share spectral shapes, and almost every feature is present in each spectrum. This is a strong indicator that measurements made using the O-PTIR technique are consistent and repeatable. Even with extreme similarity, there are slight differences in the spectra shown. 

Differences are often the shape of the peaks. Sometimes the curve is smoother or a dip is deeper and not as wide. These differences could likely be attributed to the "turbulence" of the air between the emission of the laser to the top of the sample. Additionally, the instrument has been found to be sensitive to vibrations near and on the optical table. This sensitivity could be a major contributor to the variation in spectra, which is why it is customary to use air-cushion stabilized legs for the table holding our O-PTIR instrument. The differing features will be discussed in Section \ref{s:uncertainty}. The repeatability of these measurements, even across materials, is strong evidence for the non-destructive nature of the instrument and measurement technique when power settings are adjusted correctly. 

\subsubsection{Uncertainty Analysis}
\label{s:uncertainty}

\begin{figure}[H]
    \centering
    \includegraphics[width = \textwidth]{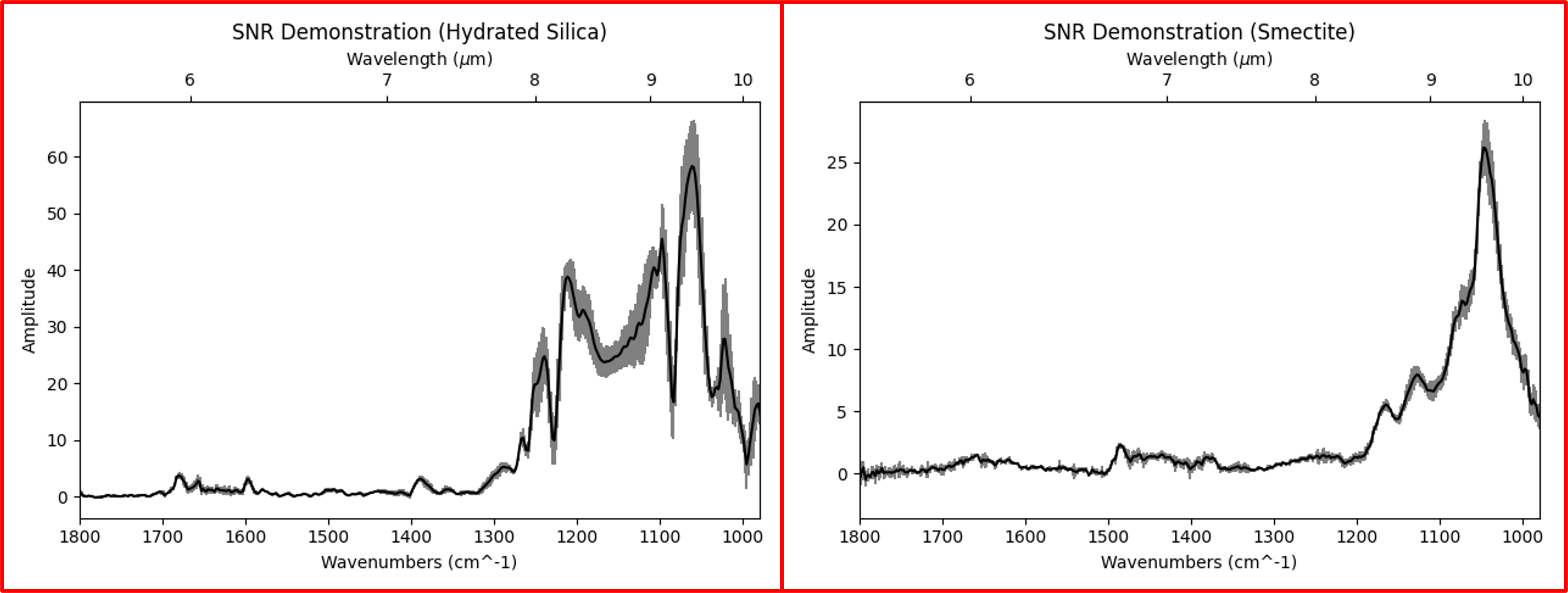}
    \caption{Mid-IR O-PTIR measurements of hydrated silica (left) and smectite (right). The measurements presented here are averaged spectral measurements of the spectra presented in Figure \ref{f:repeatability} (Section \ref{s:repeatability}). For the hydrated silica, instrument settings were: 46\% IR Power, 3.5\% Probe Power, and 5x Detector Gain. Nine spectra were averaged together for this spectrum. For the smectite, instrument settings were: 21\% IR Power, 2\% Probe Power, and 10x Detector Gain. Six spectra were averaged together for this spectrum. The averaged spectrum is presented with error bars which are representative of standard deviation.}
    \label{f:uncertainty}
\end{figure}

Figure \ref{f:uncertainty} shows an averaged spectrum of both hydrated silica (left) and smectite (right) with uncertainties plotted as error bars. These materials were chosen for their prominent features in the wavenumber range of our instrument. Hydrated silica has larger deviation than smectite, but still shows small deviation compared to its peaks. Hydrated silica had a max average peak of 58.37 with a max deviation of 10.80. That is a max error percentage of approximately 18\%. Smectite has a max average peak of 26.17 with a max deviation of 2.64. That is a max error percentage of approximately 10\%. This also shows that deviation varies by material.

\section{FTIR Comparison}
\label{s:ftir-comp}

\begin{figure}[H]
    \centering
    \includegraphics[width=\textwidth]{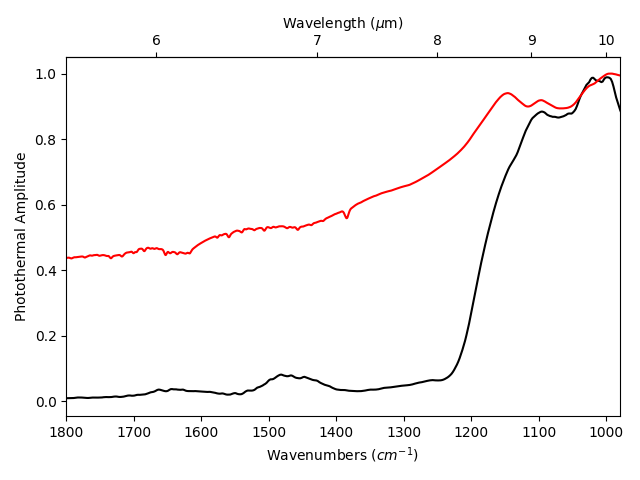}
    \caption{A comparison of O-PTIR (black) and FTIR (red) measurements of the same sample of anorthosite. The O-PTIR and FTIR measurements are both normalized to 1 for ease of comparison. The O-PTIR measurement utilized the same instrument settings as those described in Figure \ref{f:database-comparison} and is normalized by the same factor as well (0.92). The FTIR instrument settings are the same as those described in Section \ref{s:ftir-methods} and the measurement was normalized by a factor of 6.54 (i.e. every point in the spectrum was multiplied by 6.54). The 128 scan at 4 resolution is presented here.}
    \label{f:optir-ftir-comp}
\end{figure}

Figure \ref{f:optir-ftir-comp} shows a comparison of an O-PTIR measurement and an FTIR measurement of the same anorthosite sample. To account for the lesser spatial resolution of the FTIR technique, an average of a hyperspectral map (with 3 scans at each point in the map) was used as the O-PTIR measurement. The spectral resolution of 4 for the FTIR measurement was chosen so it would be comparable to the spectral resolution in our O-PTIR measurements (see Section~\ref{s:measurements}). 

The two measurements presented here both peak in similar locations. The second FTIR peak (peaking around 1100 cm$^{-1}$) has a lesser signal strength than the first peak (around 1000 cm$^{-1}$). The third FTIR peak (peaking around 1200 cm$^{-1}$) has a signal strength between that of the first and second peaks. O-PTIR matches the first and second peaks well. However, there is no significant feature at 1200 cm$^{-1}$ in the O-PTIR measurement. O-PTIR shows a stronger signal response in this specific wavenumber range. The spectral shapes of both O-PTIR and FTIR are similar in the range where the material peaks most significantly. Both measurements show a decrease in signal following 1250 wavenumbers but the signal strength for O-PTIR dropped off more significantly than the FTIR.

\begin{table}[htbp]
  \centering
  \caption{Table of comparisons between the O-PTIR and FTIR techniques. Both techniques performed 3 single spectral measurements with various numbers of scans for each technique. These measurements were performed on the same sample of anorthosite. The times reported are the time required for the respective techniques and instruments to complete the measurements. Times are reported as m:ss. Max peak is a report of the max signal measured in the wavenumber range reported. Max deviation is a report of the largest standard deviation measured in the wavenumber range reported.}
    \begin{tabular}{|l|r|r|r|r|r|r|r|r|}
    \toprule
    \rowcolor[rgb]{ 0,  0,  0}       & \multicolumn{8}{c|}{\cellcolor[rgb]{ 1,  1,  1}O-PTIR} \\
    \midrule
    Scans & 1     & 3     & 5     & 8     & 16    & 32    & 64    & 128 \\
    \midrule
    Time  & 0:09  & 0:16  & 0:23  & 0:33  & 1:02  & 1:58  & 3:52  & 7:38 \\
    \midrule
    Max Peak & 0.74  & 0.77  & 0.73  & 0.62  & 0.58  & 0.46  & 0.43  & 0.36 \\
    \midrule
    Max Deviation & 0.14  & 0.18  & 0.12  & 0.11  & 0.04  & 0.06  & 0.03  & 0.04 \\
    \midrule
    \rowcolor[rgb]{ 0,  0,  0}       & \multicolumn{8}{c|}{\cellcolor[rgb]{ 1,  1,  1}FTIR} \\
    \midrule
    Scans & 1     & 3     & 5     & 8     & 16    & 32    & 64    & 128 \\
    \midrule
    Time  & 0:01  & 0:02  & 0:04  & 0:07  & 0:14  & 0:28  & 0:56  & 1:52 \\
    \midrule
    Max Peak & 0.15  & 0.15  & 0.15  & 0.15  & 0.15  & 0.15  & 0.15  & 0.15 \\
    \midrule
    Max Deviation & 0.00  & 0.00  & 0.00  & 0.00  & 0.00  & 0.00  & 0.00  & 0.00 \\
    \bottomrule
    \end{tabular}%
  \label{t:optir-ftir-comp}%
\end{table}%

Table \ref{t:optir-ftir-comp} presents a direct comparison of O-PTIR and FTIR measurement requirements and outputs at similar spectral resolutions. It shows no deviation in the FTIR measurements because the deviation between the three measurements of each set of scans was less than .001 in all cases. Additionally, the signal was more consistent in the FTIR technique reporting approximately 0.15 for each set of scans. However, there was greater signal strength, though higher deviation, seen in the same range in the O-PTIR measurements. 

Though the FTIR technique performed scans faster and with more precision, it is important to emphasize that this is only reporting the time required for scans. The time required for measurements is far longer. In the case of FTIR, a sample must be ground to a specific grain size, a pellet must be made, and the instrument must be purged. For making a pellet, there may be several attempts required as the pellet may have imperfections rendering it useless. Additionally, pellets may not yield enough signal and must again be recreated. For the purging, often many hours are required. In our case, 12 hours of purging was required limiting us to one set of measurements per day without the ability to analyze multiple samples. This purging is required to achieve the signal needed for a valid measurement. 

O-PTIR measurements did not require any form of purging and significantly less preparation than an FTIR measurement. The measurement similarities are seen in Figure \ref{f:optir-ftir-comp}. In our case, the only preparation needed for these measurements was to fill a cylindrical container with granular samples possessing a wide range of grain sizes. In the case of monolithic samples, the sample is simply placed in the instrument with no preparation. Due to the absence of these requirements discussed for FTIR, many materials and samples can be and were analyzed in one day and in a significantly shorter amount of time.

\section{Discussion}
\label{s:discuss}

In this section, we will discuss the attributes that make this technique unique and how it offers advancements to modern planetary science missions. Additionally, we will go in depth on the benefits of the technology utilizing some of the discussion from Section \ref{s:characterization}. These benefits will be discussed through the lens of how they benefit planetary science and future planetary science space missions.

\subsection{High Resolution}
\label{s:high-res}

One advantage of O-PTIR technology is the increased spatial and spectral resolution compared to traditional IR methods. That is not to say that the technique's capability in either spatial or spectral alone is impressive, but the combination of the two together. The data presented in this work benefited by having single spectral measurements with a spatial resolution on the sub-micron scale. Traditional spatial resolution of mid-IR measurements is often limited to 10-20 microns due to the diffraction limit of the IR beam. 

The spatial resolution of O-PTIR is only limited by the diffraction limit of the probe laser. This allowed us to more definitively say what was being measured. In a planetary science mission, this would allow for the disentangling of mineral mixtures, the identification of amino acids, and the study of cellular structures. This would be an advancement in modern mission instrumentation. 

In addition to improved spatial resolution, the measurements presented in this work also benefited from having increased spectral resolution. That is to say with a spectral range of 860 wavenumbers (1820 – 960 cm$^{-1}$ in the case of our instrument), there are 860 spectral channels. Though this range of channels is commonly available in FTIR instrumentation, the level of spectral resolution at the speed at which the data is achieved is what makes it an improvement. This is discussed in Section \ref{s:ftir-comp}. 

Additionally, the ability to scan as this instrument does with the spectral resolution it has would be an advancement in technology of planetary science missions. The technique allows for the expansion of this spectral range and an increase in channels. Here, the limitation is the number of chips in the QCL being used to produce the photothermal effect. This would be a reproducible and cost-effective way to achieve data collection during missions.

\subsection{Fast Measurements}
\label{s:fast-meas}

Measurements collected using O-PTIR technology can be performed extraordinarily fast. A single hyperspectral measurement took approximately 15 s. This measurement could be performed faster with a small increase in uncertainty. A hyperspectral map with over 1000 points (three hyperspectral spectra averaged into one spectrum per point) could be completed in less than 10 hours. More poignantly, this is approximately 4 hyperspectral measurements per minute including time to average each set of individual spectra and moving the objective to the next point. We can determine the limiting factor in this case is the hardware of the instrument. This time could be improved upon given faster motors to move the sample stage or better processors recording and averaging the recorded spectra. In the case of planetary science missions, this scanning could be utilized to quickly scan areas (via an arm attachment for instance). This would serve as a fast preliminary method for searching for specific organics or inorganics.

Table \ref{t:optir-ftir-comp} presents the capabilities of O-PTIR compared to traditional FTIR-A methods. The comparison is also discussed in Section \ref{s:ftir-comp}. Specifically, it compares resolution and measurement speed. The data presented shows the ability of the instrument to collect high-quality data in a reduced amount of time compared to traditional methods. This does not only include increased spatial and spectral resolution as better discussed above, but also includes less uncertainty than traditional methods collected at the same time. This improvement would present a strong advantage for inclusion into future planetary science missions for in-situ grain or surface analysis.

\subsection{No Sample Preparation}
\label{s:no-prep}

Though we did prepare samples, as described above in section \ref{s:sample-prep}, the instrument and technology did not require it. Samples were prepared to reduce dust contamination of the optics and sample area of the instrument. The lack of necessity for sample preparation is demonstrated by the quality measurements produced with the relative lack of preparation. As described in Section \ref{s:sample-prep}, the only preparation performed was the pouring of grains into a container and then leveling the surface. 

Spectroscopy methods discussed in Section \ref{s:spectroscopy-intro} require sample preparation. This preparation can often be difficult, destructive, and/or time consuming. Time consuming in this section strictly refers to the time it takes for the sample preparation proportion and does not include the time needed for a measurement. Some FTIR methods utilize thin slices for solid samples. Disadvantages to this method include the difficulty of preparing a slice of a sample thin enough for light to pass through it as well as the destruction required to prepare the sample (via the cutting). 

For granular materials, the pellet method is generally used. Sometimes monolithic samples are ground down to make granular samples for the purposes of using this method. Turning a monolithic sample into a granular sample is of course a destructive process. The pellet method itself is not destructive though it is difficult to complete. Pellets often are too brittle to use or some other complication (such as cloudy or cracked pellets) prevents the pellet from being analyzed. 

Ices can also be analyzed using FTIR. There are several ways to do this, but some methods rely on condensing the ice directly onto a sample holder in an ultra-high vacuum. This may not be damaging, but it is certainly difficult and time consuming. This brief description of FTIR sample preparation is not all inclusive but does convey the difficulty. In comparison, IR absorption of sample surface ices can simply and rapidly be achieved using O-PTIR.

O-PTIR not requiring sample preparation is an outstanding quality. The simplicity of placing a sample in the instrument with no preparation is in stark contrast with other state-of-the-art methods. In-situ planetary science missions would greatly benefit from the lack of need to prepare samples. It would reduce hardware complexity as the only requirement on the sample is to be approachable by an objective. In the laboratory, countless hours would be saved being able to simply collect a sample and begin to process it. Experimentalists would greatly benefit from not needing to utilize man hours or other resources preparing samples for analysis. 

It was mentioned in Section \ref{s:optir-intro} that in some configurations of this spectrometry method, Raman is used simultaneously. Simultaneous Raman could aid in the unambiguous identification of minerals and amino acids. This is shown by \cite{krafft2022optical}. The addition of Raman to O-PTIR would aid in the additional identification features still with no sample preparation. However, this is out of the scope of this work.

\section{Conclusion and Summary}
\label{s:conclusion}

In this work, we presented O-PTIR spectra of various minerals and demonstrated the capabilities of the O-PTIR technique. We demonstrated the relevance of the method as it is comparable to existing data (this will be shown further in a future work). We also make a case for the use of this technique in the planetary science community. The benefits of high spatial and spectral resolution, fast measurements, and the reduced need for sample preparation were demonstrated and discussed. We also discussed applications to future space missions.

In future work, we will present the ability of O-PTIR to analyze and characterize materials relevant to planetary science. In doing so we will also make more comparisons of the O-PTIR measurements of materials to FTIR absorbance measurements of materials. Further, we will demonstrate the capability to perform quantitative analysis on opaque materials.

\bibliography{literature}{}
\bibliographystyle{elsarticle-harv}

\end{document}